\documentstyle[11pt]{article} 
\input epsf

 \newcommand{\be}{\begin{equation}}
 \newcommand{\ee}{\end{equation}}
\newcommand{\bea}{\begin{eqnarray}} \newcommand{\eea}{\end{eqnarray}}
\begin{document}

\begin{center}
{\LARGE{\bf Exact Local Bosonic Algorithm for Full QCD}}\\[1cm]
{\small Proceeding of the Workshop on {\em Accelerating Fermion Algorithms} at HLRZ J\"ulich (Germany), February 21st-23rd, 1996}\\[1cm]

\vspace*{1cm}
{\Large Ph. de Forcrand$^{a,}$\footnote{forcrand@scsc.ethz.ch}, A. Galli$^{b,}$\footnote{galli@mppmu.mpg.de}}\\[0.3cm]
$^a${\small \em SCSC,ETH-Zentrum, CH-8092 Z\"urich, Switzerland}\\
$^b${\small \em Max-Planck-Institut f\"ur Physik, D-80805 Munich, Germany}
\vspace*{1cm}
\end{center}
\begin{abstract}
We present an exact local bosonic algorithm for the simulation of
dynamical fermions in lattice QCD. It is based on a non-hermitian 
polynomial approximation of the inverse of the quark matrix and a  
global Metropolis accept/reject correction of the systematic errors. 
We show that this algorithm is a real alternative to the Hybrid Monte Carlo 
algorithm.
\end{abstract}

\section{Introduction}

The search for more efficient full QCD algorithms has motivated substantial
activity, both within the classical Hybrid Monte Carlo (HMC) and in
the alternative method of the local bosonic algorithm.
This last algorithm was proposed by L\"uscher in a hermitian version 
\cite{luescher1}.
The idea was to approximate the full QCD partition function using a local
bosonic action based on a polynomial approximation of the inverse of the
squared Wilson fermion matrix. The aim was to obtain an algorithm which
does not need the explicit inversion of the fermion matrix and which
only uses local, finite-step-size updates, contrary to HMC.\\

In this work we study several significant improvements of 
this algorithm:  an inexpensive stochastic Metropolis accept/reject test
\cite{forcrand1} to make the algorithm exact;  a non-hermitian polynomial approximation \cite{forcrand1} and
 a simple even-odd preconditioning \cite{beat1,ultimo}. 
Here we consider two-flavor QCD, and we present a description of the 
algorithm. We illustrate the 
effectiveness of our algorithm with representative Monte Carlo results. We 
successfully model its static properties (the Monte Carlo acceptance), and try 
to disentangle its dynamics.\\

From our analysis it appears that this version of the local bosonic algorithm 
is a real alternative to the classic HMC.
For heavy quarks its finite-step dynamics are comparable to 
quenched Monte Carlo, and are considerably faster than HMC.
And for large volumes and small quark masses, the scaling of our algorithm
compares favorably with that of HMC. 
Thus our algorithm becomes attractive
for light quarks on large lattices as well. 
Finally, because it uses local updating techniques,
it is not affected by the accumulation of roundoff errors which can marr
the reversibility of HMC in such cases.

\section{Description of the algorithm}

The full QCD partition function with two fermion flavors is given by
\be Z=\int [dU] |\det D|^2 e^{-S_G[U]}
\label{fullQCD}
\ee where $D$ represents the fermion matrix and $S_G$ denotes the pure gauge action. This partition function can be approximated introducing a polynomial
approximation of the inverse of the fermion matrix.  \be |\det
D|^2=\det D^\dagger\det D\simeq \frac{1}{|\det P(D)|^2}
\label{app}
\ee The polynomial $P(z)=\prod_{k=1}^n(z-z_k)$ of degree $n$ is
defined in the complex plane and approximates the inverse of $z$. 
Since we are investigating full QCD with two flavors, the determinant 
$|\det P(D)|^2$ manifestly factorizes into positive pairs, so that the
$\frac{1}{|\det P(D)|^2}$ 
term of the approximation (\ref{app}) can be expressed
by a Gaussian integral over a set of boson fields $\phi_k$ ($k=1,...,n$)
with color and Dirac indices. The full
QCD partition function (\ref{fullQCD}) is then approximated by 
\be
Z\simeq \int [dU][d\phi][d\phi^\dagger]\ e^{-S_L[U,\phi]}
\label{lus}
\ee 
where $\phi$ represents the set of all boson field families,
$S_L$ is the local action $S_L=S_G+S_b$ and 
$
S_b=\sum_{k=1}^n 
|(D-z_k)\phi_k|^2
$.
Making use of the locality of $S_L$ 
we may now simulate the partition function (\ref{lus}) by locally updating
the boson fields and the gauge fields, using heat-bath and
over-relaxation algorithms. \\ 

The simulation of full QCD can be obtained from (\ref{lus}) by correcting
the errors due to the approximation through a Metropolis test at the end of each trajectory.
Introducing the error term in the partition function we obtain 
\be 
Z=\int [dU] [d\phi] [d\phi^\dagger]|\det(DP(D))|^2e^{-S_L(U,\phi)}\label{exact}
\ee

The correction term $ |\det(DP(D))|^2$ can be evaluated in two different 
ways.\\

The first one consists in estimating the determinant $ |\det(DP(D))|^2$
using a noisy estimator \cite{ultimo}.
The strategy of this method is to update the $(U,\phi)$ fields such that the probability of finding a particular configuration is proportional to $e^{-S_L}$ and then perform a Metropolis test defining an acceptance probability $P^A$ in 
terms of the noisy estimation of the correction $|\det(DP(D))|^2$.
The acceptance probability $P^A$ is given by 
\be
P^A_{(U,\phi)\rightarrow(U',\phi')}=
\min\left(1,e^{-|W\chi|^2+|\chi|^2}\right)
\label{32}
\ee
where $W=[D'P(D')]^{-1}DP(D)$. In this case the algorithm satisfies detailed balance after averaging over the Gaussian noise $\chi$.\\

The second method consists in expressing the correction term $ |\det(DP(D))|^2$
directly by a Gaussian integral and incorporating the dynamics of the correction field $\eta$ in the partition sum (\ref{exact})
\be
Z=\int [dU][d\eta][d\eta^\dagger]
 [d\phi] [d\phi^\dagger]e^{-S_{exact}(U,\phi,\eta)}
\label{exact2}
\ee
by defining a new exact action
\be
S_{exact}(U,\phi,\eta)=S_{L}(U,\phi)+S_{C}(U,\eta)
\ee
where 
$
S_{C}(U,\eta)=|[DP(D)]^{-1}\eta|^2
$
is the correction action. In carrying out the simulation one generates 
configurations of $(U,\phi)$ and $\eta$ such that the probability of 
finding a particular configuration is proportional to $\exp(-S_{exact})$. 
Also in this case the strategy is to alternatively update the $(U,\phi)$ 
fields and the $\eta$ fields. The Metropolis acceptance  probability 
$P^A$ is not defined using a noisy estimation of the correction 
$ |\det(DP(D))|^2$, but directly using the exact action $S_{exact}$ 
so that the transition probability of the algorithm 
satisfies detailed balance {\em without} the need to average over $\eta$.
In this case the acceptance probability is again defined in terms of a
Gaussian vector $\chi$, by
\be
P^A_{(U,\phi)\rightarrow(U',\phi')}=
\left\{\begin{array}{lr}
\min\left(1,e^{-|W\chi|^2+|\chi|^2}\right)&\mbox{ if } S_G(U)\geq S_G(U')\\
\min\left(1,e^{+|W^{-1}\chi|^2-|\chi|^2}\right)&\mbox{ if } S_G(U)<S_G(U')
\end{array}
\right.\label{df}
\ee

We have tested both methods and both seem equally efficient. In \cite{ultimo} we have presented a formal proof that both algorithms converge to the right distribution.\\

The algorithm can be summarized as follows:
\begin{itemize}
\item Generate a new Gaussian spinor $\chi$ with variance one.
\item Update locally the boson and gauge fields $m$ times (in reversible order)
according to the approximate partition function (\ref{lus}).
\item Accept/reject the new configuration according to the Metropolis acceptance
probability (\ref{32}) for the noisy version 
or (\ref{df}) for the non-noisy version. 
\end{itemize}

In order to evaluate the Metropolis acceptance probability we have to solve 
a linear system involving $DP(D)$ or $D'P(D')$, 
for which we use the BiCGstab algorithm
 \cite{schilling}. This linear system is very well conditioned because $P(D')$ 
(or $P(D)$) approximates the inverse of $D'$ (or $D$). 
The cost for solving it is minimal
and scales like the local updating algorithms in the volume and quark mass. \\

We emphasize that our algorithm remains exact for any choice of the polynomial
$P$. 
If the polynomial approximates the inverse of the fermion matrix 
well, the acceptance of the Metropolis correction will be high; 
if not the acceptance will be low. 
The number and location of the roots $z_k$ in the complex plane determine the 
quality of the approximation and hence the acceptance. 
Since the algorithm is exact for any polynomial, a priori 
knowledge about the spectrum of the fermion matrix is not required.

\section{Results for the exact local bosonic algorithm.}

Numerical simulations using the exact local bosonic algorithm in the non-noisy version described in the previous section have been performed  for different 
lattice parameters.
The majority of the simulations are reported in \cite{ultimo}.\\

We explored the acceptance of the Metropolis correction test and the number of 
iterations of the BiCGStab algorithm used in that test for inverting $DP(D)$.   
The study was performed by varying the degree $n$ of the polynomial and the hopping parameter $k$. 
One observes that the acceptance increases quite rapidly with the degree of 
the polynomial, above some threshold (see Figs. 1,2,3). On the other hand, the number of 
iterations needed to invert $DP(D)$ remains very low for high enough acceptance. The data show clearly that the overhead due to the Metropolis test remains 
negligible provided that the degree of the polynomial is tuned to have 
sufficient acceptance. Results obtained from simulations using 
even-odd preconditioning confirm, as expected, that the improvement of the 
approximation 
reduces the required number of bosonic fields by at least a factor two.

\section{Predicting the Metropolis acceptance}

Using some general assumptions and the known error bounds on the Chebyshev-like 
approximation of $D^{-1}$, we can obtain \cite{ultimo} an ansatz
for the acceptance probability, which at $\beta = 0$ takes the form
\begin{equation}
< P_{acc} > \simeq erfc\left( f \sqrt{96 V} 
\left(\frac{K}{K_c}\right)^{n + 1} \right)
\label{fitacc}
\end{equation}
where $V$ is the lattice volume, $K$ is the hopping parameter of the Wilson fermion matrix $D$ and $K_c$ is the critical hopping parameter.
We expect the fitting parameter
$f$ to depend smoothly on $\beta$, but very little on $n, V$ and $K$.
Figs. 1, 2 and 3 show the acceptance we measured during our Monte-Carlo
simulations at $\beta = 0$, as a function of $n$, for 2 different volumes 
and 2 different $K$'s. The fit (eq. (\ref{fitacc})) is shown by the dotted
lines. All 3 figures have been obtained with the same value $f =0.19$.
We thus consider our ansatz (\ref{fitacc}) quite satisfactory.
At other values of $\beta$, one can fix $f$ by a test on a small lattice,
and then predict the acceptance for larger volumes and different quark masses.
We will use our ansatz below to analyze the cost of our algorithm.

\section{Understanding the dynamics}

        The coupled dynamics of the gauge and bosonic fields in L\"uscher's
method are rather subtle. The long autocorrelation times 
$\tau\propto n$ were explained in \cite{beat1,ultimo,borrelli} for the local algorithm {\em without} the global Metropolis test. \\

Since we now have a reasonable understanding of the Metropolis acceptance and
of the dynamics without the Metropolis test, we can see the effect of the one
on the other, and then estimate the total cost of the algorithm per independent
configuration. Calling $\tau$
and $\tau_0$ the autocorrelation time with and without Metropolis respectively,
we obtain \cite{ultimo}
\begin{equation}
\tau = \frac{-1}{\log(1 - <P_{acc}> (1 - e^{-m/\tau_0}))}
\label{tau}
\end{equation}
for a trajectory of $m$ sweeps. Note that $\tau_0$ is measured in sweeps and 
$\tau$ in trajectories. Folding into (\ref{tau}) our ansatz for $<P_{acc}>$ (eq. (\ref{fitacc})),
and measuring the proportionality constant $C$ for the autocorrelation without the Metropolis test $\tau_0 \sim Cn$, we obtain  
the autocorrelation time as a function of $n$. An example is plotted in Fig. 4 for lattices of sizes 4, 8, 16, 32, at
$\beta = 0$ and $K = 0.215$, with $m = 10$ (for this lattice parameters $C\sim 1.3)$. The behavior will be qualitatively
similar for other choices of parameters. The Monte Carlo data shown in Fig. 4 was
obtained on a $4^4$ lattice, and is roughly compatible with eq.(\ref{tau}).\\

The operation count of the algorithm is about $6 n$ matrix-vector multiplications
by the Dirac operator $D$ per sweep. Therefore we can estimate the total cost of
our algorithm to produce an independent configuration, measured in multiplications
by $D$ per lattice site, as a function of the number of fields or the Metropolis
acceptance. As expected, one can observe that the optimal number of bosonic
fields grows logarithmically with the volume; 
the acceptance must be kept high, with a fairly broad optimum around $70 - 80\%$.

\section{Scaling}

The scaling of the number of bosonic fields and of the total complexity with
the volume and the quark mass has already been discussed in \cite{forcrand1,beat1,ultimo,borrelli}.
Our analysis confirms these earlier estimates.\\

$\bullet$ As the volume $V$ increases, the number $n$ of bosonic fields should
grow like $log V$. 
This is a consequence of the exponential convergence of the
polynomial approximation. Since the work per
sweep is proportional to $n V$, and the autocorrelation time to $n$, the work
per independent configuration grows like $V (\log V)^2$.
This is an asymptotically slower growth than Hybrid Monte Carlo which requires work $\sim V^{5/4}$.
But this is more of an academic than a practical advantage.\\

$\bullet$ As the quark mass $m_q$ decreases, the number of fields $n$ must grow like
$m_q^{-1}$. This can be derived from (\ref{fitacc}), using
$m_q \propto 1 - K/K_c$, which applies for small quark masses. 
The work per sweep is proportional to $n$. The autocorrelation time 
behavior is less clear. We expect that the autocorrelation behaves like
$\tau \sim n m_q^{-\alpha}$, since $m_q$ enters in the effective mass term
of each harmonic piece of the action $S_b$, and since
a factor $n$ comes from the autocorrelation of the gauge fields alone. 
There is no theoretical understanding of the coupling between the gauge and boson fields,
so to determine $\alpha$ our main 
argument comes from MC data. From an exploratory simulation \cite{ultimo} on a $4^4$ 
lattice at $\beta=0$ we obtained 
that $\alpha$ is near $1$. This is only 
indicative, because the dependence of $\alpha$ with the volume and the 
coupling has yet to be explored. \\

\section{Conclusion}

We have presented an alternative algorithm to HMC for simulating dynamical 
quarks in lattice QCD. This algorithm is based on a local bosonic action. 
A non-hermitian polynomial approximation of the inverse of the quark matrix 
is used to define the local bosonic action. The addition of a global 
Metropolis test corrects the systematic errors. The overhead of the correction 
test is minimal. Even-odd preconditioning is very simple to implement. It 
reduces the number of required bosonic fields by at least a factor two, and
accelerates the dynamics by the same factor.\\

This algorithm is exact for any choice of the polynomial approximation. 
No critical tuning of the approximation parameters is needed. Only the efficiency
of the algorithm, which can be monitored, will be affected by the choice of parameters.
The cost of the algorithm increases with the volume $V$ of the lattice as $V(\log V)^2$
and with the inverse of the quark mass $m_q$ as $\frac{n^2}{m_q^\alpha}\sim 
\frac{1}{m_q^{2+\alpha}}$ with  
$\alpha\simeq 1$. 
This compares favorably with the scaling of HMC. \\

Finally, for heavy quarks the dynamic properties of our algorithm approach 
those of quenched Monte Carlo, and are considerably faster than HMC in that
regime \cite{rajan}. 
We thus have presented an algorithm superior to HMC in the limits of
heavy and light dynamical quarks, and we expect it to be competitive in the
intermediate regime.

\begin{figure}
\centerline{\epsfysize = 8 in \epsffile {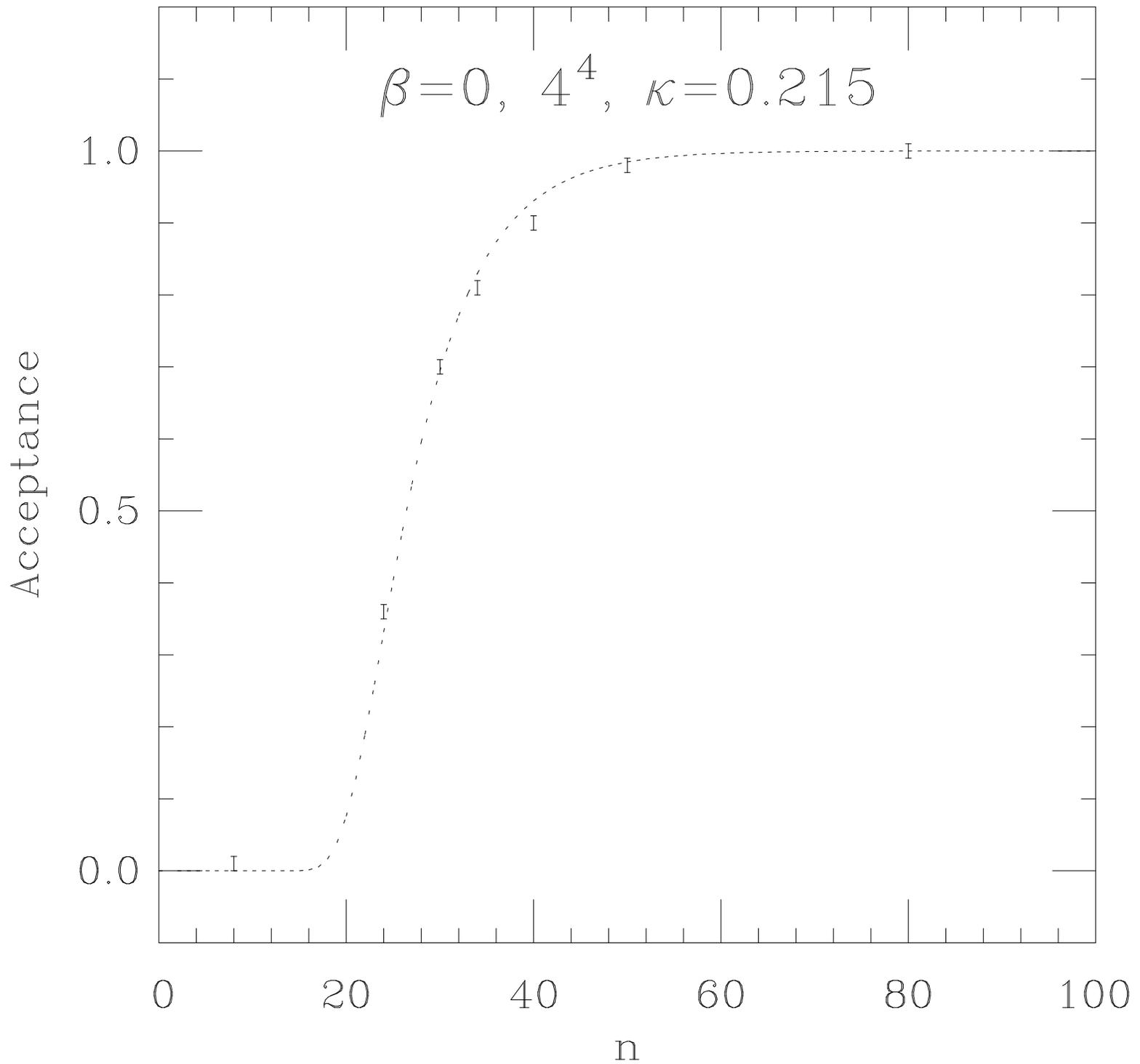}}
\caption{ Metropolis acceptance as a function of the number of bosonic
fields. The dotted line is our 1-parameter ansatz eq.(9).}
\end{figure}

\begin{figure}
\centerline{\epsfysize = 8 in \epsffile {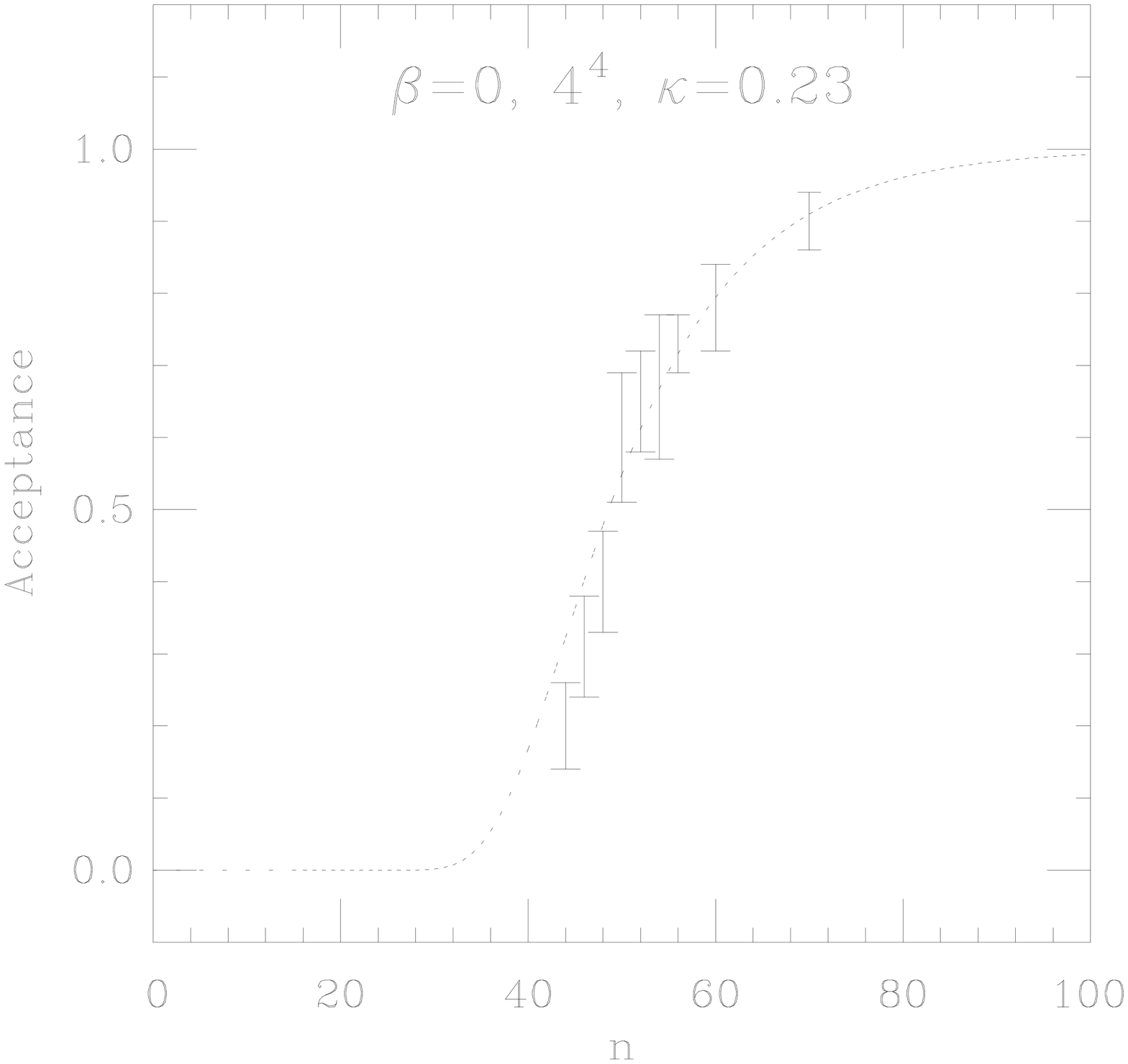}}
\caption{ Same as Fig.1, for a different value of $K$. The fit
parameter $f$ (eq.(9)) is unchanged from Fig.1.}
\end{figure}

\begin{figure}
\centerline{\epsfysize = 8 in \epsffile {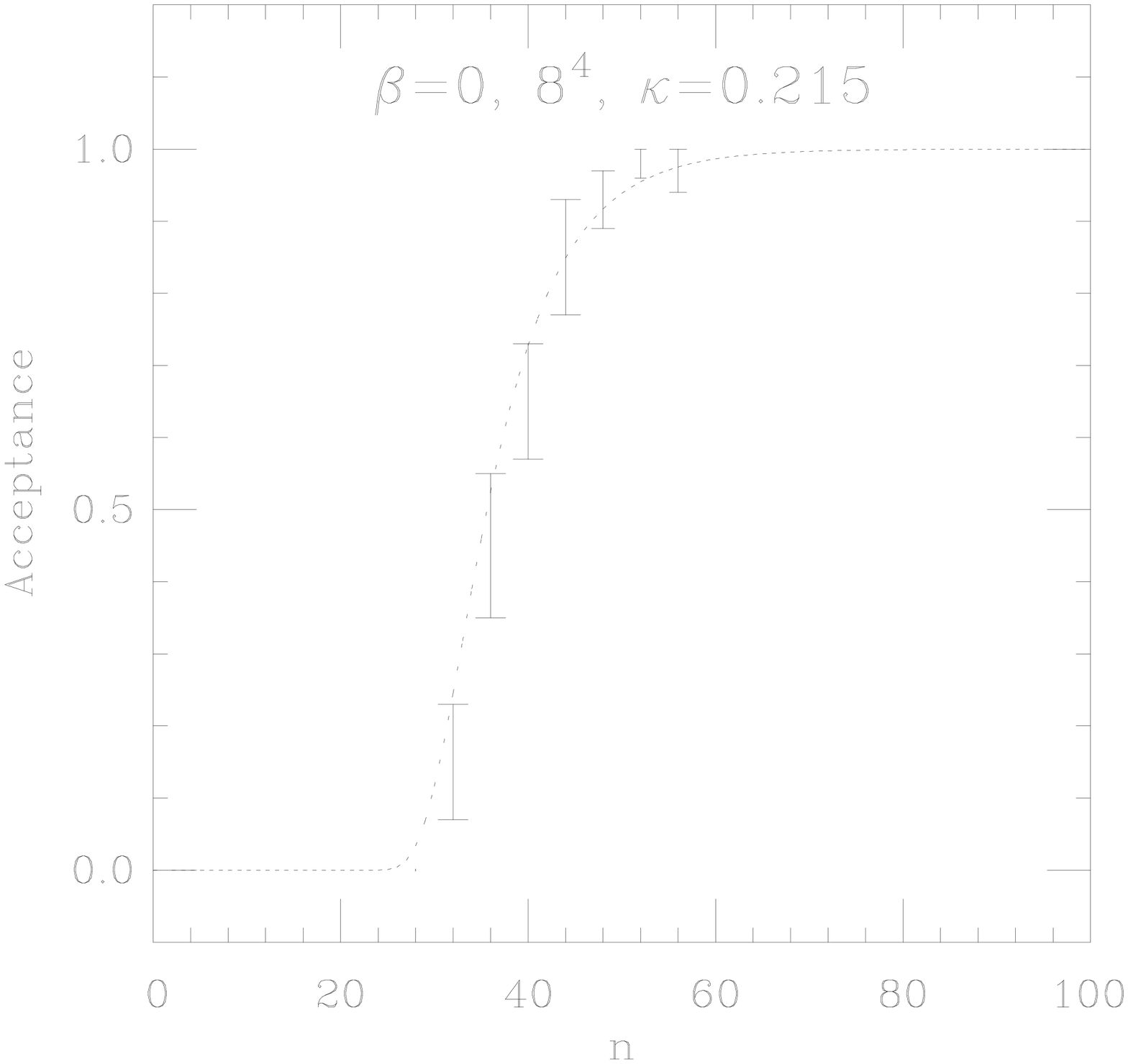}}
\caption{ Same as Fig.1, for a different volume. The fit parameter $f$
(eq.(9)) is unchanged from Fig.1.}
\end{figure}

\begin{figure}
\centerline{\epsfysize = 8 in \epsffile {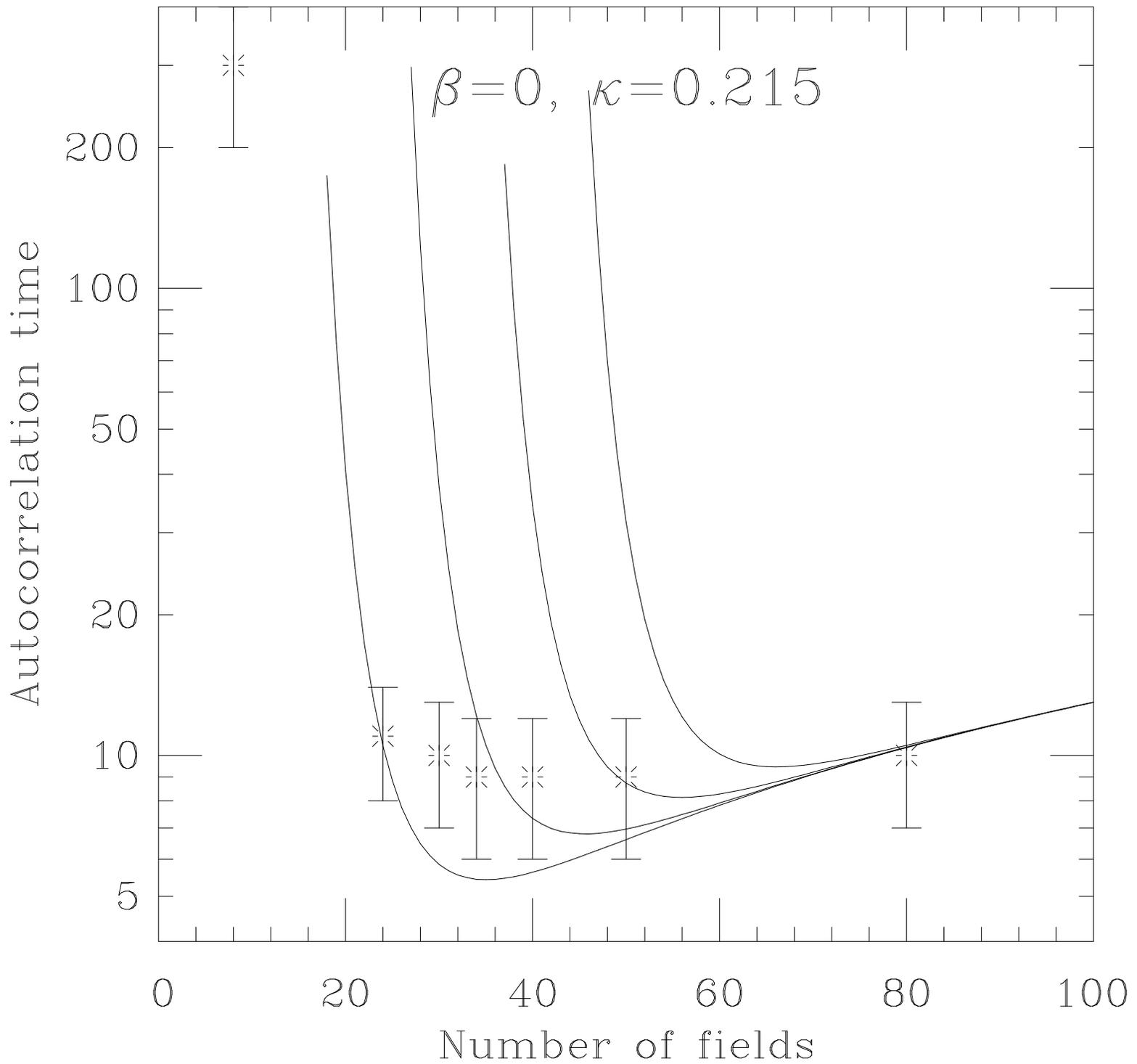}}
\caption{Integrated autocorrelation time given by our ansatz (10)
as a function of the number of fields, measured in trajectories, for lattices
of size $L = 4, 8, 16, 32$ from left to right. The Monte Carlo results also
shown have been obtained for $L = 4$.}
\end{figure}

\end{document}